\documentclass[acmsmall]{acmart}

\AtBeginDocument{%
  \providecommand\BibTeX{{%
    \normalfont B\kern-0.5em{\scshape i\kern-0.25em b}\kern-0.8em\TeX}}}

\copyrightyear{2019}
\acmYear{2019}

\acmConference[KDD 2019 Workshop]{KDD 2019: Workshop on Applied Data Science for Healthcare}{August 05, 2019}{Anchorage, AK, USA}

\hypersetup{
    colorlinks=true,
    linkcolor=blue,
    filecolor=magenta,      
    urlcolor=cyan,
}

\begin{document}

\title[BCKG]{An Information Extraction and Knowledge Graph Platform for Accelerating Biochemical Discoveries}

\author{Matteo Manica}
\orcid{0000-0002-8872-0269}
\email{tte@zurich.ibm.com}
\author{Christoph Auer}
\email{cau@zurich.ibm.com}
\author{Valery Weber}
\email{vwe@zurich.ibm.com}
\author{Federico Zipoli}
\email{fzi@zurich.ibm.com}
\author{Michele Dolfi}
\email{dol@zurich.ibm.com}
\author{Peter Staar}
\email{taa@zurich.ibm.com}
\author{Teodoro Laino}
\email{teo@zurich.ibm.com}
\author{Costas Bekas}
\email{bek@zurich.ibm.com}
\affiliation{%
  \institution{IBM Research}
  \streetaddress{Säumerstrasse 4}
  \city{Rüschlikon}
  \state{Zürich}
  \postcode{8803}
  \country{Switzerland}}
\email{tte@zurich.ibm.com}
\author{Akihiro Fujita}
\email{akihiro.fujita@hb.nagase.co.jp}
\affiliation{%
  \institution{HAYASHIBARA \& CO., LTD.}
  \country{Japan}}
\author{Hiroki Toda}
\email{hiroki.toda@nagase.co.jp}
\author{Shuichi Hirose}
\email{shuichi.hirose@nagase.co.jp}
\author{Yasumitsu Orii}
\email{yasumitsu.orii@nagase.co.jp}
\affiliation{%
  \institution{NAGASE \& CO., LTD.}
  \country{Japan}}

\renewcommand{\shortauthors}{Manica et al.}
\newcommand{\tte}[1]{\textcolor{purple}{#1}}

\begin{abstract}
  Information extraction and data mining in biochemical literature is a daunting task that demands resource-intensive computation and appropriate means to scale knowledge ingestion.
  Being able to leverage this immense source of technical information helps to drastically reduce costs and time to solution in multiple application fields from food safety to pharmaceutics.
  We present a scalable document ingestion system that integrates data from databases and publications (in PDF format) in a biochemistry knowledge graph (BCKG).
  The BCKG is a comprehensive source of knowledge that can be queried to retrieve known biochemical facts and to generate novel insights.
  After describing the knowledge ingestion framework, we showcase an application of our system in the field of carbohydrate enzymes.
  The BCKG represents a way to scale knowledge ingestion and automatically exploit prior knowledge to accelerate discovery in biochemical sciences.
\end{abstract}




\maketitle

\section{Introduction}

\begin{figure}[!htb]
  \centering
  \includegraphics[width=\textwidth]{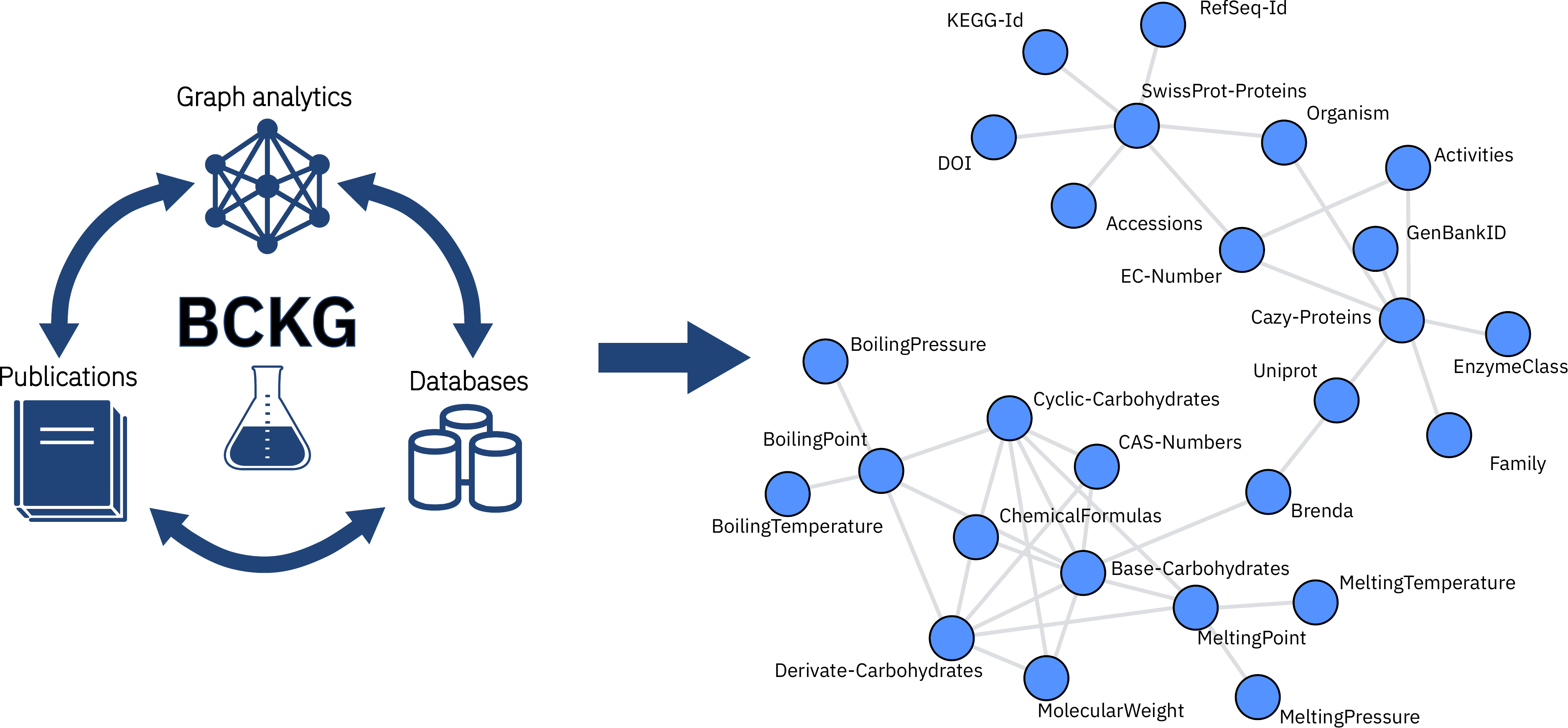}
\begin{small}
	\caption{\textbf{BCKG concept and current structure.} The platform ingests knowledge from different data sources and implements graph analytics techniques in a comprehensive and queryable knowledge base (left). The currently assembled KG integrates multiple data sources organizing them in linked collections of nodes (right).
	\label{fig:overview}
	}
\end{small}
\Description{BCKG - concept and current structure.}
\end{figure}

The discovery of novel biotechnological processes to produce chemicals and materials with competitive industrial conditions is of paramount importance for new and old businesses. Relying on opportunistic or occasional discoveries prevents innovation that could be pursued instead, by reasoning on the large knowledge corpus of science collected in the last century. For this reason, the industrial competitive advantage of the next decade will be strongly connected with the possibility to extract and represent the immense human knowledge accumulated in the past to accelerate the discovery of new processes and materials.
To get a glimpse of the task's complexity, as of writing, by simply querying NCBI~\citep{NCBIResourceCoordinators2016} for the keyword \texttt{carbohydrate} we retrieve: >220K genes, >1.5M papers, >12.9M proteins.
Ideally, we would like to be able to ingest all this information at scale and translate it right away into actionable insights.
For example, in the context of food or pharmaceutical ingredients, carbohydrates play a crucial role and it is of primary importance to retrieve information about any characterized enzyme able to synthesize them.
The design of a digital platform implementing the possibility to explore the vast amount of data available will speed up and improve the discovery of novel ingredients and methodologies to synthesize them, directly impacting key aspects of the health care and life sciences domain, e.g., food safety, drug design.
Over the years efforts to collect biomolecular information in a structured format have led to the compilation of multiple database resources: protein-specific databases, enzyme-specific databases, chemical compound databases and various resources regarding pathways, taxonomy information, DNA or protein sequences.
While these resources are extremely valuable, they present two problematic aspects: first, despite of the fact that some databases integrate multiple sources or report links to other knowledge bases, having independent data collections hinders our ability to effectively reason on knowledge, find patterns or generate novel insights; secondly, a large portion of scientific and technical knowledge is stored in an unstructured format in publications or in books.
Recently, an interesting framework, called \texttt{biochem4j}, has been developed by~\citet{Swainston2017}.
This resource integrates multiple databases with a focus on metabolic engineering in a queryable knowledge graph (KG).
While extremely useful, it only addresses one of the aforementioned issues, as it overlooks the information contained in publications and books in the form of natural language, tables, figures, etc.
To bridge this gap, herein we present a scalable document ingestion platform for biochemical knowledge: BCKG (see~\autoref{fig:overview}). The platform integrates information from multiple database resources and it leverages recent developments in machine learning-based document parsing to assemble a comprehensive biochemistry Knowledge Graph (KG). Organizing knowledge in a graph structure allows to integrate disparate data resources and to efficiently retrieve the knowledge ingested.
Moreover, graph analytic techniques enable the generation of novel insights, a key aspect in research and industrial applications.

\section{Methods}

\subsection{Data Ingestion}

The BioChemistry Knowledge Graph (BCKG) is created from two distinct, separate sources. On the one hand, we ingest data from a comprehensive list of structured databases: UniProt~\citep{TheUniProtConsortium2018}, Pfam~\citep{El-Gebali2019}, BRENDA~\citep{Jeske2019}, CAZy~\citep{Cantarel2009}, PubChem~\citep{Kim2019}, ChEMBL~\citep{Gaulton2017}, KEGG~\citep{Kanehisa2000,Kanehisa2017}, The NCBI Taxonomy database~\citep{Federhen2012}, GenBank~\citep{Clark2016} and PDB~\citep{Berman2000}.
On the other hand, we also ingest unstructured data, such as the  Handbook of Carbohydrate Engineering~\citep{Yarema2005} and scientific articles.
The ingestion of both type of sources into a single KG presents unique challenges.
First, the size of some databases is large and it is therefore necessary to have a scalable, cloud-native platform to host this data. Furthermore, the data within the databases is not represented in a unique way (e.g., the concept of the EC number is represented in at least 3 different ways), which requires us to \textit{normalize} all database structures in order to link key-value pairs from one database to another. Second, the ingestion of scientific articles and books in (scanned) PDF format poses a real problem, since the PDF format does not allow one to easily extract the information encoded in the documents. To address this issue we use the Corpus Conversion Service (CCS) \cite{Staar2018}, a scalable, cloud-native platform to convert large collections of PDF documents into a structured JSON format. The latter contains all the titles, abstracts, section-titles, paragraphs, tables, images in textual form. As such, it easily allows us to perform Named Entity Recognition (NER) and Fact Extraction (FE) on the text/tables of the original documents. Using NER, we can find the entities referred to in the databases (e.g., taxonomies, genes, proteins, etc). Using the FE, we can find properties of these entities in the documents.   

\subsection{KG construction}

The KG is assembled by executing the following steps consecutively. First, we parse the structured databases representing each entity in a JSON document with normalized keys. Second, we extract and aggregate common concepts from these documents (e.g., the EC number) and create links between the extracted items. These links form indirect connections between the documents obtained from the database. Next, we parse the handbooks and the scientific articles using the CCS. The latter provides us with a JSON file for each PDF-document. From these JSON documents, we then extract paragraphs and tables, on which we perform NER and FE. The NER provides us with entities found in the text/tables and allows us to link structured databases with the text/tables from the PDF-documents via co-occurrence. Furthermore, we also aggregate all facts form the FE process, which allows us to further add links between entities (e.g., organisms producing carbohydrates with similar properties). The resulting KG contains >6M nodes and >61M edges.

\subsection{KG queries and analytics}

Our platform supports a wide range of queries and standard graph analytics: node retrieval, graph traversal, centrality analysis, clustering, etc.
A unique feature of the BCKG resides in its query-engine, capable of running long workflows on the graph.
A workflow is a group of queries linked in a Directed Acyclic Graph (DAG). Given their DAG structure, workflows can be used to efficiently perform complex queries, allowing us to explore the graph and to potentially generate novel insights.

\section{Results and discussion}

To showcase the BCKG platform, we implemented a KG workflow to retrieve enzymes synthesizing Trehalose, a carbohydrate that has major applications in food, cosmetics and pharmaceutics~\cite{Higashiyama2002}.
Its extraction has been optimized around 2000 by the Hayashibara Company (NAGASE Group) discovering a cost-effective biotech process using starch as a starting material.
The search of new enzymes producing Trehalose is a field of research still very active and with a great business potential.
For carbohydrate research, CAZy represents the relevant source of knowledge. In an effort to plan for serendipity, we compiled a query to identify, in the larger UniProt database (Swiss-Prot, manually reviewed), candidate enzymes able to process Trehalose that have not yet been reported in CAZy. 
The resulting workflow is composed of:
\begin{enumerate}
\item Trehalose node identification in the Handbook of Carbohydrate Engineering.
\item Graph traversal in the UniProt database to gather all proteins whose catalytic activity node has a direct connection with the Trehalose node.
\item Identify those UniProt protein nodes that have no connection in CAZy.
\end{enumerate}
Among the several hits returned by this query, we report: (a) stf0 - Trehalose 2-sulfotransferase - Mycobacterium tuberculosis (strain ATCC 25618 / H37Rv) and (b) TPP1 - Probable trehalose-phosphate phosphatase 1 - Oryza sativa subsp. japonica (Rice). The TPP1 is quite interesting as the phosphatase class is well represented in CAZy and still this specific enzyme was not listed.

This example, in its simplicity, demonstrates the great capabilities of knowledge graphs in planning for serendipity, accelerating the discovery process thanks to the effective representation of the domain knowledge. 

\bibliographystyle{ACM-Reference-Format-Truncation}
\bibliography{main}


\begin{thebibliography}{16}


\ifx \showCODEN    \undefined \def \showCODEN     #1{\unskip}     \fi
\ifx \showDOI      \undefined \def \showDOI       #1{#1}\fi
\ifx \showISBNx    \undefined \def \showISBNx     #1{\unskip}     \fi
\ifx \showISBNxiii \undefined \def \showISBNxiii  #1{\unskip}     \fi
\ifx \showISSN     \undefined \def \showISSN      #1{\unskip}     \fi
\ifx \showLCCN     \undefined \def \showLCCN      #1{\unskip}     \fi
\ifx \shownote     \undefined \def \shownote      #1{#1}          \fi
\ifx \showarticletitle \undefined \def \showarticletitle #1{#1}   \fi
\ifx \showURL      \undefined \def \showURL       {\relax}        \fi
\providecommand\bibfield[2]{#2}
\providecommand\bibinfo[2]{#2}
\providecommand\natexlab[1]{#1}
\providecommand\showeprint[2][]{arXiv:#2}

\bibitem[\protect\citeauthoryear{Benson, Cavanaugh, Clark,
  et~al\mbox{.}}{Benson et~al\mbox{.}}{2017}]%
        {Clark2016}
\bibfield{author}{\bibinfo{person}{Dennis~A. Benson}, \bibinfo{person}{Mark
  Cavanaugh}, \bibinfo{person}{Karen Clark}, {et~al\mbox{.}}}
  \bibinfo{year}{2017}\natexlab{}.
\newblock \showarticletitle{{GenBank}}.
\newblock \bibinfo{journal}{\emph{Nucleic Acids Research}}
  \bibinfo{volume}{45}, \bibinfo{number}{D1} (\bibinfo{date}{jan}
  \bibinfo{year}{2017}), \bibinfo{pages}{D37--D42}.
\newblock
\showISSN{13624962}
\urldef\tempurl%
\url{https://doi.org/10.1093/nar/gkw1070}
\showDOI{\tempurl}


\bibitem[\protect\citeauthoryear{Berman, Battistuz, Bhat, et~al\mbox{.}}{Berman
  et~al\mbox{.}}{2002}]%
        {Berman2000}
\bibfield{author}{\bibinfo{person}{Helen~M. Berman}, \bibinfo{person}{Tammy
  Battistuz}, \bibinfo{person}{T.~N. Bhat}, {et~al\mbox{.}}}
  \bibinfo{year}{2002}\natexlab{}.
\newblock \showarticletitle{{The protein data bank}}.
\newblock \bibinfo{journal}{\emph{Acta Crystallographica Section D: Biological
  Crystallography}} \bibinfo{volume}{58}, \bibinfo{number}{6 I}
  (\bibinfo{date}{jan} \bibinfo{year}{2002}), \bibinfo{pages}{899--907}.
\newblock
\showISSN{09074449}
\urldef\tempurl%
\url{https://doi.org/10.1107/S0907444902003451}
\showDOI{\tempurl}


\bibitem[\protect\citeauthoryear{Cantarel, Coutinho, Rancurel,
  et~al\mbox{.}}{Cantarel et~al\mbox{.}}{2009}]%
        {Cantarel2009}
\bibfield{author}{\bibinfo{person}{Brandi~I. Cantarel},
  \bibinfo{person}{Pedro~M. Coutinho}, \bibinfo{person}{Corinne Rancurel},
  {et~al\mbox{.}}} \bibinfo{year}{2009}\natexlab{}.
\newblock \showarticletitle{{The Carbohydrate-Active EnZymes database (CAZy):
  An expert resource for glycogenomics}}.
\newblock \bibinfo{journal}{\emph{Nucleic Acids Research}}
  \bibinfo{volume}{37}, \bibinfo{number}{SUPPL. 1} (\bibinfo{date}{jan}
  \bibinfo{year}{2009}), \bibinfo{pages}{D233--8}.
\newblock
\showISSN{03051048}
\urldef\tempurl%
\url{https://doi.org/10.1093/nar/gkn663}
\showDOI{\tempurl}


\bibitem[\protect\citeauthoryear{El-Gebali, Mistry, Bateman,
  et~al\mbox{.}}{El-Gebali et~al\mbox{.}}{2019}]%
        {El-Gebali2019}
\bibfield{author}{\bibinfo{person}{Sara El-Gebali}, \bibinfo{person}{Jaina
  Mistry}, \bibinfo{person}{Alex Bateman}, {et~al\mbox{.}}}
  \bibinfo{year}{2019}\natexlab{}.
\newblock \showarticletitle{{The Pfam protein families database in 2019}}.
\newblock \bibinfo{journal}{\emph{Nucleic Acids Research}}
  \bibinfo{volume}{47}, \bibinfo{number}{D1} (\bibinfo{date}{jan}
  \bibinfo{year}{2019}), \bibinfo{pages}{D427--D432}.
\newblock
\showISSN{13624962}
\urldef\tempurl%
\url{https://doi.org/10.1093/nar/gky995}
\showDOI{\tempurl}


\bibitem[\protect\citeauthoryear{Federhen}{Federhen}{2012}]%
        {Federhen2012}
\bibfield{author}{\bibinfo{person}{Scott Federhen}.}
  \bibinfo{year}{2012}\natexlab{}.
\newblock \showarticletitle{{The NCBI Taxonomy database}}.
\newblock \bibinfo{journal}{\emph{Nucleic Acids Research}}
  \bibinfo{volume}{40}, \bibinfo{number}{D1} (\bibinfo{date}{jan}
  \bibinfo{year}{2012}), \bibinfo{pages}{D136--43}.
\newblock
\showISSN{03051048}
\urldef\tempurl%
\url{https://doi.org/10.1093/nar/gkr1178}
\showDOI{\tempurl}


\bibitem[\protect\citeauthoryear{Gaulton, Hersey, Nowotka,
  et~al\mbox{.}}{Gaulton et~al\mbox{.}}{2017}]%
        {Gaulton2017}
\bibfield{author}{\bibinfo{person}{Anna Gaulton}, \bibinfo{person}{Anne
  Hersey}, \bibinfo{person}{Micha~L. Nowotka}, {et~al\mbox{.}}}
  \bibinfo{year}{2017}\natexlab{}.
\newblock \showarticletitle{{The ChEMBL database in 2017}}.
\newblock \bibinfo{journal}{\emph{Nucleic Acids Research}}
  \bibinfo{volume}{45}, \bibinfo{number}{D1} (\bibinfo{year}{2017}),
  \bibinfo{pages}{D945--D954}.
\newblock
\showISSN{13624962}
\urldef\tempurl%
\url{https://doi.org/10.1093/nar/gkw1074}
\showDOI{\tempurl}


\bibitem[\protect\citeauthoryear{Higashiyama}{Higashiyama}{2002}]%
        {Higashiyama2002}
\bibfield{author}{\bibinfo{person}{Takanobu Higashiyama}.}
  \bibinfo{year}{2002}\natexlab{}.
\newblock \showarticletitle{{Novel functions and applications of trehalose}}.
\newblock \bibinfo{journal}{\emph{Pure and Applied Chemistry}}
  \bibinfo{volume}{74}, \bibinfo{number}{7} (\bibinfo{date}{jan}
  \bibinfo{year}{2002}), \bibinfo{pages}{1263--1269}.
\newblock
\showISSN{0033-4545}
\urldef\tempurl%
\url{https://doi.org/10.1351/pac200274071263}
\showDOI{\tempurl}


\bibitem[\protect\citeauthoryear{Jeske, Placzek, Schomburg,
  et~al\mbox{.}}{Jeske et~al\mbox{.}}{2019}]%
        {Jeske2019}
\bibfield{author}{\bibinfo{person}{Lisa Jeske}, \bibinfo{person}{Sandra
  Placzek}, \bibinfo{person}{Ida Schomburg}, {et~al\mbox{.}}}
  \bibinfo{year}{2019}\natexlab{}.
\newblock \showarticletitle{{BRENDA in 2019: A European ELIXIR core data
  resource}}.
\newblock \bibinfo{journal}{\emph{Nucleic Acids Research}}
  \bibinfo{volume}{47}, \bibinfo{number}{D1} (\bibinfo{date}{jan}
  \bibinfo{year}{2019}), \bibinfo{pages}{D542--D549}.
\newblock
\showISSN{13624962}
\urldef\tempurl%
\url{https://doi.org/10.1093/nar/gky1048}
\showDOI{\tempurl}


\bibitem[\protect\citeauthoryear{Kanehisa, Furumichi, Tanabe,
  et~al\mbox{.}}{Kanehisa et~al\mbox{.}}{2017}]%
        {Kanehisa2017}
\bibfield{author}{\bibinfo{person}{Minoru Kanehisa}, \bibinfo{person}{Miho
  Furumichi}, \bibinfo{person}{Mao Tanabe}, {et~al\mbox{.}}}
  \bibinfo{year}{2017}\natexlab{}.
\newblock \showarticletitle{{KEGG: New perspectives on genomes, pathways,
  diseases and drugs}}.
\newblock \bibinfo{journal}{\emph{Nucleic Acids Research}}
  \bibinfo{volume}{45}, \bibinfo{number}{D1} (\bibinfo{date}{jan}
  \bibinfo{year}{2017}), \bibinfo{pages}{D353--D361}.
\newblock
\showISSN{13624962}
\urldef\tempurl%
\url{https://doi.org/10.1093/nar/gkw1092}
\showDOI{\tempurl}


\bibitem[\protect\citeauthoryear{Kim, Chen, Cheng, et~al\mbox{.}}{Kim
  et~al\mbox{.}}{2019}]%
        {Kim2019}
\bibfield{author}{\bibinfo{person}{Sunghwan Kim}, \bibinfo{person}{Jie Chen},
  \bibinfo{person}{Tiejun Cheng}, {et~al\mbox{.}}}
  \bibinfo{year}{2019}\natexlab{}.
\newblock \showarticletitle{{PubChem 2019 update: Improved access to chemical
  data}}.
\newblock \bibinfo{journal}{\emph{Nucleic Acids Research}}
  \bibinfo{volume}{47}, \bibinfo{number}{D1} (\bibinfo{date}{jan}
  \bibinfo{year}{2019}), \bibinfo{pages}{D1102--D1109}.
\newblock
\showISSN{13624962}
\urldef\tempurl%
\url{https://doi.org/10.1093/nar/gky1033}
\showDOI{\tempurl}


\bibitem[\protect\citeauthoryear{Ogata, Goto, Sato, et~al\mbox{.}}{Ogata
  et~al\mbox{.}}{1999}]%
        {Kanehisa2000}
\bibfield{author}{\bibinfo{person}{Hiroyuki Ogata}, \bibinfo{person}{Susumu
  Goto}, \bibinfo{person}{Kazushige Sato}, {et~al\mbox{.}}}
  \bibinfo{year}{1999}\natexlab{}.
\newblock \showarticletitle{{KEGG: Kyoto encyclopedia of genes and genomes}}.
\newblock \bibinfo{journal}{\emph{Nucleic Acids Research}}
  \bibinfo{volume}{27}, \bibinfo{number}{1} (\bibinfo{date}{jan}
  \bibinfo{year}{1999}), \bibinfo{pages}{29--34}.
\newblock
\showISSN{03051048}
\urldef\tempurl%
\url{https://doi.org/10.1093/nar/27.1.29}
\showDOI{\tempurl}


\bibitem[\protect\citeauthoryear{Sayers, Agarwala, Bolton,
  et~al\mbox{.}}{Sayers et~al\mbox{.}}{2019}]%
        {NCBIResourceCoordinators2016}
\bibfield{author}{\bibinfo{person}{Eric~W. Sayers}, \bibinfo{person}{Richa
  Agarwala}, \bibinfo{person}{Evan~E. Bolton}, {et~al\mbox{.}}}
  \bibinfo{year}{2019}\natexlab{}.
\newblock \showarticletitle{{Database resources of the National Center for
  Biotechnology Information}}.
\newblock \bibinfo{journal}{\emph{Nucleic Acids Research}}
  \bibinfo{volume}{47}, \bibinfo{number}{D1} (\bibinfo{date}{jan}
  \bibinfo{year}{2019}), \bibinfo{pages}{D23--D28}.
\newblock
\showISSN{13624962}
\urldef\tempurl%
\url{https://doi.org/10.1093/nar/gky1069}
\showDOI{\tempurl}


\bibitem[\protect\citeauthoryear{Staar, Dolfi, Auer, et~al\mbox{.}}{Staar
  et~al\mbox{.}}{2018}]%
        {Staar2018}
\bibfield{author}{\bibinfo{person}{Peter W~J Staar}, \bibinfo{person}{Michele
  Dolfi}, \bibinfo{person}{Christoph Auer}, {et~al\mbox{.}}}
  \bibinfo{year}{2018}\natexlab{}.
\newblock \showarticletitle{{Corpus Conversion Service}}. In
  \bibinfo{booktitle}{\emph{Proceedings of the 24th ACM SIGKDD International
  Conference on Knowledge Discovery {\&} Data Mining - KDD '18}}.
  \bibinfo{publisher}{ACM Press}, \bibinfo{address}{New York, New York, USA},
  \bibinfo{pages}{774--782}.
\newblock
\showISBNx{9781450355520}
\urldef\tempurl%
\url{https://doi.org/10.1145/3219819.3219834}
\showDOI{\tempurl}


\bibitem[\protect\citeauthoryear{Swainston, Batista-Navarro, Carbonell,
  et~al\mbox{.}}{Swainston et~al\mbox{.}}{2017}]%
        {Swainston2017}
\bibfield{author}{\bibinfo{person}{Neil Swainston}, \bibinfo{person}{Riza
  Batista-Navarro}, \bibinfo{person}{Pablo Carbonell}, {et~al\mbox{.}}}
  \bibinfo{year}{2017}\natexlab{}.
\newblock \showarticletitle{{biochem4j: Integrated and extensible biochemical
  knowledge through graph databases}}.
\newblock \bibinfo{journal}{\emph{PLoS ONE}} \bibinfo{volume}{12},
  \bibinfo{number}{7} (\bibinfo{date}{jul} \bibinfo{year}{2017}),
  \bibinfo{pages}{e0179130}.
\newblock
\showISSN{19326203}
\urldef\tempurl%
\url{https://doi.org/10.1371/journal.pone.0179130}
\showDOI{\tempurl}


\bibitem[\protect\citeauthoryear{{The UniProt Consortium}}{{The UniProt
  Consortium}}{2018}]%
        {TheUniProtConsortium2018}
\bibfield{author}{\bibinfo{person}{{The UniProt Consortium}}.}
  \bibinfo{year}{2018}\natexlab{}.
\newblock \showarticletitle{{UniProt: a worldwide hub of protein knowledge}}.
\newblock \bibinfo{journal}{\emph{Nucleic Acids Research}}
  \bibinfo{volume}{47}, \bibinfo{number}{D1} (\bibinfo{date}{jan}
  \bibinfo{year}{2018}), \bibinfo{pages}{D506--D515}.
\newblock
\showISSN{0305-1048}
\urldef\tempurl%
\url{https://doi.org/10.1093/nar/gky1049}
\showDOI{\tempurl}


\bibitem[\protect\citeauthoryear{Yarema}{Yarema}{2010}]%
        {Yarema2005}
\bibfield{author}{\bibinfo{person}{Kevin~J. Yarema}.}
  \bibinfo{year}{2010}\natexlab{}.
\newblock \bibinfo{booktitle}{\emph{{Handbook of Carbohydrate Engineering}}}.
\newblock \bibinfo{publisher}{Taylor {\&} Francis}. 904 pages.
\newblock
\showISBNx{9781574444728}
\urldef\tempurl%
\url{https://doi.org/10.1201/9781420027631}
\showDOI{\tempurl}


\end{thebibliography}


\end{document}